%% file: mainHjj.tex
\documentclass[preprint,twocolumn,5p,nopreprintline,sort&compress]{elsarticle}
\usepackage{epsfig}
\usepackage{amsbsy,amssymb} 
\usepackage{graphicx}
\usepackage{array}
\usepackage{amsmath}
\usepackage{feynarts}
\usepackage{multirow}
\newcommand\new{\newcommand}         

\def\beq{\begin{equation}}   
\def\eeq{\end{equation}}
\def\bea{\begin{eqnarray}}  
\def\eea{\end{eqnarray}} 
\newcommand{\bite}{\begin{itemize}}
\newcommand{\eite}{\end{itemize}}

\def\gev{\; \mathrm{GeV}}

\new{\eV}         {{\ifmmode {\mathrm{ eV}}\else ${\mathrm{ eV}}$\fi}}
\new{\MeV}        {{\ifmmode {\mathrm{ MeV}}\else ${\mathrm{ MeV}}$\fi}}
\new{\MeVc}       {{\ifmmode {\mathrm{ MeV}}/c\else ${\mathrm{ MeV}}/c$\fi}}
\new{\MeVcc}      {{\ifmmode {\mathrm{ MeV}}/c^2\else ${\mathrm{ MeV}}/c^2$\fi}}
\new{\GeV}        {{\ifmmode {\mathrm{ GeV}}\else ${\mathrm{ GeV}}$\fi}}
\new{\GeVc}       {{\ifmmode {\mathrm{ GeV}}/c\else ${\mathrm{GeV}}/c$\fi}}
\new{\GeVcc}      {{\ifmmode {\mathrm{ GeV}}/c^2\else ${\mathrm{GeV}}/c^2$\fi}}
\new{\TeV}        {{\ifmmode {\mathrm{ TeV}}\else ${\mathrm{ TeV}}$\fi}}

\new{\Mh}         {{\ifmmode M_{\mathrm{ H}}
                    \else $M_{\mathrm{H}}$\fi}}

\new{\Mz}         {{\ifmmode M_{\mathrm{Z}}
                    \else $M_{\mathrm{Z}}$\fi}}
\new{\Mzsq}       {{\ifmmode M^2_{\mathrm{ Z}}
                    \else $M^2_{\mathrm{Z}}$\fi}}

\new{\as}[1]      {{\ifmmode\alpha^{#1}_s
                    \else$\alpha^{#1}_s$\fi}}
\new{\asx}[1]      {{\ifmmode a^{#1}_s
                    \else $a^{#1}_s$\fi}}
\new{\asb}[1]     {{\ifmmode\overline{\alpha}^{#1}_s
                    \else $\overline{\alpha}^{#1}_s$\fi}}
\new{\asmz}       {{\ifmmode\alpha_s(\Mzsq)
                    \else $\alpha_s(\Mzsq)$\fi}}
\new{\lqcd}       {{\ifmmode\Lambda_{\mathrm{ QCD}}
                    \else $\Lambda_{\mathrm{ QCD}}$\fi}}

\def\Feynarts{{{\sc FeynArts}}}
\def\Gosam{{{\sc GoSam}}}

\def\samurai{{{\sc samurai}}}
\def\Sherpa{{{\sc Sherpa}}}

\def\C++{{{\sc c++}}}
\def\Powheg{{{\sc Powheg}}}
\def\MCFM{{{\sc MCFM}}}
\def\QCDLoop{{{\sc QCDLoop}}}
\def\OneLoop{{{\sc OneLoop}}}
\def\Golem{{{\sc Golem95C}}}
\def\FastJet{{{\sc FastJet}}}

\newcommand{\ci}{g_{\mbox{\tiny eff}}}

\newcommand{\mpi}{Max-Planck-Institut f\"ur Physik, F\"ohringer Ring 6, 80805 M\"unchen, Germany}
\newcommand{\padova}{Dipartimento di Fisica e Astronomia, Universit\`a di Padova, and INFN  \\                                    
Sezione di Padova, via Marzolo 8, 35131 Padova, Italy}
\newcommand{\cuny}{New York City College of Technology, City University of New York, 300 Jay Street, Brooklyn NY 11201, USA}
\newcommand{\cunygc}{The Graduate School and University Center, City University of New York, 365 Fifth Avenue, New York, NY 10016, USA}
\newcommand{\nuni}{Dipartimento di Scienze Fisiche, Universit\`a degli studi di Napoli ``Federico II'', I-80125 Napoli, Italy}
\newcommand{\ninfn}{INFN, Sezione di Napoli, I-80125 Napoli, Italy  }

\begin{document}

\begin{frontmatter}


\title{NLO QCD corrections to the production of Higgs  plus two jets  at the LHC
\tnoteref{t1}}

\tnotetext[t1]{DSF-03-2013, MPP-2013-36}
\address[mpi]{\mpi}
\address[pad]{\padova}
\address[cuny]{\cuny}
\address[cunygc]{\cunygc}
\address[nuni]{\nuni}
\address[ninfn]{\ninfn}

\author[mpi]{H. van Deurzen}
\ead{hdeurzen@mpp.mpg.de}
%
\author[mpi]{N. Greiner}
\ead{greiner@mpp.mpg.de}
\author[mpi]{G. Luisoni} 
\ead{luisonig@mpp.mpg.de}
%
\author[mpi,pad]{P. Mastrolia}
\ead{pierpaolo.mastrolia@cern.ch}  
%
\author[mpi]{E. Mirabella}
\ead{mirabell@mpp.mpg.de}  
%
%
\author[cuny,cunygc]{G. Ossola} 
\ead{GOssola@citytech.cuny.edu}
%
\author[mpi]{T. Peraro}
\ead{peraro@mpp.mpg.de}
%
%
\author[mpi]{J.~F.~von~Soden-Fraunhofen} 
\ead{jfsoden@mpp.mpg.de}
%
%
\author[nuni,ninfn]{F. Tramontano}
\ead{francesco.tramontano@cern.ch}
%

\begin{abstract}
We present the calculation of the NLO QCD corrections to the associated
production of a Higgs boson and two jets,
in the infinite top-mass limit.  We discuss the technical details of the computation
and we show the numerical impact of the radiative corrections on
several observables at the LHC. 
The results are obtained by using a fully automated framework for
fixed order NLO QCD calculations based on the interplay of the packages 
\Gosam\ and  \Sherpa .
The evaluation of the virtual corrections constitutes an application of the 
$d$-dimensional integrand-level reduction to theories with higher dimensional operators.
We also present first results for the one-loop matrix elements  
of the partonic processes with a quark-pair in the final state, which
enter the hadronic production of a Higgs boson together with three
jets in the infinite top-mass approximation.
\end{abstract}


\end{frontmatter}



\section{Introduction}
\label{Sec:intro}

The recent discovery reported by ATLAS and CMS~\cite{:2012gk,:2012gu}
indicates the existence of a new neutral boson with mass of about $125$ {\GeV}
and  spin different from one. At present, all measurements are consistent with
the hypothesis that the new particle is the Standard Model Higgs boson.
Nevertheless, further studies concerning its CP properties, spin, and
couplings 
are mandatory to confirm its nature.

The vector boson fusion (VBF) processes can be used to study the CP properties
of the new particle, and to extract its couplings with the heavy gauge bosons.
The dominant background to VBF comes from Higgs plus two jet production via gluon
fusion ($Hjj$). The signal-over-background ratio can be improved by
imposing stringent cuts on the Higgs decay products, and by requiring large rapidity
separation between the two forward jets.
A veto on the jet activity
in the central region further reduces the impact of the background. Indeed,
 VBF processes are characterized by  low hadronic activity, owing to the  exchange of a
color-singlet in the $t$-channel. An accurate estimation of  the efficiency of the central-jet veto requires
the inclusion of the  next-to-leading order (NLO) corrections
to $Hjj$ production~\cite{Campbell:2006xx,Campbell:2010cz}.

The leading order (LO) contribution to $Hjj$ production has been
computed in Refs.~\cite{DelDuca:2001eu,DelDuca:2001fn}. 
The calculation was performed retaining the full top-mass
dependence, and it showed the validity of the large top-mass
approximation ($m_{t} \to \infty$) whenever
the mass of the Higgs particle and the $p_T$ of the jets are 
not sensibly larger than the mass of the top quark.
In the $m_{t} \to \infty$ limit, the Higgs coupling to two gluons, which at LO is mediated by a top-quark loop, becomes independent of $m_t$. Hence, it can be described
by an effective operator obtained by integrating out the top quark
degrees of freedom \cite{Wilczek:1977zn}.
In this approximation, the number of loops of the virtual diagrams
that need to be computed is reduced by one.

In the heavy top quark limit, the inclusive Higgs boson production cross section has
been computed at NLO \cite{Dawson:1990zj, Djouadi:1991tka}
and at next-to-next-to-leading order (NNLO)
\cite{Harlander:2002wh,Anastasiou:2002yz,Ravindran:2003um},
showing a reduced sensitivity of the perturbative prediction to scale variations.

The implementation of final state cuts, to reduce the impact of the
Standard Model background for the identification of Higgs production
demands fully exclusive calculations of the theory predictions.
In the $m_{t} \to \infty$ limit, the NNLO corrections to Higgs production via gluon fusion have
been computed fully exclusively~\cite{Anastasiou:2005qj,Anastasiou:2007mz,Catani:2007vq,Grazzini:2008tf}.
These corrections also included the contributions of $H + 1j$ final
states to NLO~\cite{Schmidt:1997wr, deFlorian:1999zd, Ravindran:2002dc, Glosser:2002gm}, and of the $H + 2j$ final states to LO~\cite{Dawson:1991au, Kauffman:1996ix}.

The impact of the NLO QCD corrections to the $Hjj$
production rate has been studied in Ref.~\cite{Campbell:2006xx},
using the real corrections presented in Refs.~\cite{DelDuca:2004wt, Dixon:2004za,Badger:2004ty} and the semi-analytic virtual corrections computed in Refs.~\cite{Ellis:2005qe,Ellis:2005zh}.  A great effort has been
devoted to the analytic computation of the one-loop helicity amplitudes involving Higgs plus four partons, using on-shell and generalized
unitarity methods~\cite{Berger:2006sh,Badger:2006us,Badger:2007si,Glover:2008ffa,Badger:2009hw,Dixon:2009uk,Badger:2009vh}.  The resulting compact expressions have
been implemented in \MCFM\  \cite{Campbell:2010cz}, which has been used to obtain matched NLO plus shower
predictions within the \Powheg \  box framework~\cite{Campbell:2012am}.

In this letter we present an independent computation of
the NLO contributions to Higgs plus  two jets
production at the LHC in the large top-mass limit.
These results have been obtained by using a fully automated framework for fixed order
NLO QCD calculations, which interfaces via the Binoth Les Houches
Accord (BLHA)~\cite{Binoth:2010xt}
the {\Gosam} package~\cite{Cullen:2011ac}, for the
generation and computation of the virtual amplitudes, with the {\Sherpa} package~\cite{Gleisberg:2008ta},
for the computation of the real amplitudes and the Monte Carlo integration over phase space.
Details of the {\Gosam} -{\Sherpa} interface, together with a selection of ready-to-use process packages,
can be found in~\cite{LST, packagesURL}.
Moreover, the evaluation of the virtual corrections for a model described by an effective Lagrangian constitutes
an application of the $d$-dimensional integrand reduction
\cite{Ossola:2006us,Ossola:2007bb,Ellis:2007br,Giele:2008ve,Mastrolia:2010nb,Mastrolia:2012an}
to theories with higher dimensional operators~\cite{Mastrolia:2012bu}. In Section~\ref{sec:calc} we present the computational setup, whereas results on $pp \to Hjj$ at the LHC are discussed in Section~\ref{sec:res}.

Finally, we explore the possibility of extending our framework to
consider the production of a Higgs boson plus three jets ($Hjjj$).
We generate codes for the virtual corrections to the partonic
processes with a quark-pair in the final state.
We show the corresponding results in Section~\ref{Sec:Hjjj}.

\ref{App:Feynman} contains the definitions of the Feynman rules for
the direct interaction between the Higgs boson and gluons via 
effective couplings. In~\ref{App:Q2}, we show a property of
the highest-rank terms of the one-loop integrands stemming from those effective rules.
We close this communication by collecting, in~\ref{App:Bench} and~\ref{App:BenchJ3},
the numerical results of the virtual matrix elements,
evaluated in non-exceptional phase space points,
for $H+2j$ and $H+3j$ final states, respectively.


\section{Computational setup}
\label{sec:calc}

We compute Higgs plus two jets production through gluon fusion in the large top-mass limit
($m_t \to \infty$). In this limit, the 
Higgs-gluon coupling is described by the effective local interaction~\cite{Wilczek:1977zn}
\bea
\mathcal{L} =-  \frac{\ci}{4} \, H \, \mbox{tr} \left (G_{\mu \nu} G^{\mu \nu} \right ) \, .
\label{Eq:EffL}
\eea
In the $\overline{\mbox{MS}}$  scheme, the  coefficient $\ci$ reads~\cite{Djouadi:1991tka, Dawson:1990zj}
 \bea
 \ci = - \frac{\alpha_s}{3 \pi v} \left ( 1 + \frac{11}{4 \pi}  \alpha_s\right )  + \mathcal{O}(\alpha_s^3)\, , 
 \eea
 in terms of the Higgs vacuum expectation value $v$. 
The operator~(\ref{Eq:EffL}) leads to new Feynman rules
involving the Higgs field and up to four gluons. They are collected in~\ref{App:Feynman}.
\medskip

Next-to-leading order corrections to cross sections require the evaluation of virtual and real emission contributions.
For the computation of the virtual corrections we use a code generated by the program package
{\Gosam}, which combines automated diagram generation and algebraic manipulation \cite{Nogueira:1991ex, Vermaseren:2000nd, Reiter:2009ts, Cullen:2010jv} 
with integrand-level reduction techniques \cite{Ossola:2006us, Ossola:2007bb, Ellis:2007br, Ossola:2008xq,  Mastrolia:2008jb, Heinrich:2010ax}.
More specifically, the virtual corrections are evaluated using the $d$-dimensional integrand-level decomposition implemented 
in the \samurai\ library~\cite{Mastrolia:2010nb}, which allows for the combined
determination of both cut-constructible and rational terms at once. Moreover, the presence of effective couplings in the 
Lagrangian requires an extended version~\cite{Mastrolia:2012bu} of the integrand-level reduction, of which the present calculation is a first application.
After the reduction, all relevant scalar (master) integrals are computed by means of {\QCDLoop}~\cite{vanOldenborgh:1990yc, Ellis:2007qk},
{\OneLoop}~\cite{vanHameren:2010cp}, or {\Golem}~\cite{Cullen:2011kv}.

For the calculation of tree-level contributions we use {\Sherpa}~\cite{Gleisberg:2008ta}, which computes the LO and the real radiation matrix elements~\cite{Krauss:2001iv}, 
regularizes the IR and collinear singularities using the Catani-Seymour dipole formalism~\cite{Gleisberg:2007md}, and carries out the 
phase space integrations as well.

The code that evaluates the virtual corrections is generated by {\Gosam} and linked to {\Sherpa} via the Binoth-Les-Houches Accord (BLHA)~\cite{Binoth:2010xt} interface. This interface allows to generate the code in a fully automated way by a system of {\rm order} and {\rm contract} files containing the amplitudes requested by {\Sherpa}. 
Furthermore, it allows for a direct communication between the
two codes at running time, when {\Sherpa} steers the integration by calling the external code which computes the virtual amplitude. A detailed description of this interface is beyond the scope of this
paper and will be presented elsewhere~\cite{LST}.

For $Hjj$ production, the partonic processes in the contract file are:
 \begin{align}
 & q \, q \, \to \, H \, q \, q  \, ,        &           &   q \, \bar q \, \to \, H \, q \, \bar q  \, , \nonumber \\
 & q \, \bar q \, \to \, H \, g  \, g  \, ,   &               &   q \, \bar q \, \to \, H \, q' \, \bar q'   \, , \nonumber \\
& q \, q' \, \to \, H \, q \, q'  \, ,             &       &   q \, g \, \to \, H \, g \, q  \, , \nonumber \\
& \bar q \, q \, \to \, H \,q \, \bar q  \, ,  &                  &   \bar q \, q' \, \to \, H \, q' \, \bar q  \, , \nonumber \\
& g \, q \, \to \, H \, g \, q  \, ,              &      &   g \, g \, \to \, H \, q \, \bar q  \, , \nonumber \\
& g \, g \, \to \, H \, g \, g  \, .              &      &  
 \label{Eq:Pproc}
 \end{align}
These processes are not independent, as they can be related by crossing and/or by relabeling.
\Gosam\  identifies and generates  the following minimal set of processes
 \begin{align}
 & g \, g \, \to \, H \, g \, g  \, ,        &           &   g \, g \, \to \, H \, q \, \bar q  \, , \nonumber \\
 & q \, \bar q \, \to \, H \, q   \, \bar q  \, ,   &               &   q \, \bar q \, \to \, H \, q' \, \bar q'   \, .
  \label{Eq:Pproc1}
  \end{align}
  The other processes  are obtained by performing the appropriate symmetry transformation.

 The ultraviolet (UV), the  infrared, and the collinear
 singularities are regularized using dimensional reduction (DRED).  UV divergences
 have been renormalized in the $\overline{\mbox{MS}}$ scheme.  
 In the case of LO [NLO] contributions we describe the running of the 
 strong coupling constant with one-loop [two-loop] accuracy, decoupling 
 the top quark from the running. 
\medskip

 The effective $Hgg$ coupling, see~\ref{App:Feynman}, leads to integrands that may exhibit 
numerators with rank $r$ larger than the number $n$ of the denominators, 
i.e. $r \le  n+1$. In general, for these cases, the parametrization of the residues at the multiple-cut
has to be extended as discussed  in Ref.~\cite{Mastrolia:2012bu}. As a consequence, 
the decomposition of any  one-loop $n$-point amplitude in terms of master integrals (MIs)  acquires new contributions,  
reading as,
\bea
\mathcal{M}^{\mbox{\tiny one-loop}}_n = \mathcal{A}_n + \delta \mathcal{A}_n \, . 
\label{Eq:M1loop}
\eea
The term $\mathcal{A}_n$ corresponds to  the standard decomposition 
for the case of a renormalizable theory ($r \le n$), 
while 
the additional contribution $\delta \mathcal{A}_n$ enters  whenever  $r \le n+1$. Their actual expressions
can be found in Eqs.~(2.16) and (6.11) of~\cite{Mastrolia:2012bu}.

The extended integrand decomposition has been implemented in the \samurai\ library.
In particular, the   coefficients multiplying the MIs appearing in $ \mathcal{A}_n$
and $\delta \mathcal{A}_n$  are  computed 
by using the discrete Fourier transform as described in 
Refs.~\cite{Mastrolia:2008jb,Mastrolia:2010nb}.

\medskip

In the case of Higgs plus jets production, higher rank numerators arise from diagrams 
where the Higgs boson is attached to a pure gluonic loop.
However, as shown in \ref{App:Q2}, the rank-$(n+1)$ terms  of an $n$-point integrand
are proportional to the loop momentum squared, $q^2$, which simplifies against a 
denominator. Therefore, they generate $(n-1)$-point integrands with rank $r=n-1$.
Consequently,  the coefficients of the MIs in  $\delta  \mathcal{A}_n$ have to vanish identically,
as explicitly verified. 
Since  $\delta  \mathcal{A}_n$  in Eq.~(\ref{Eq:M1loop}) does not play any role, the integrand reduction 
can be also performed  with the current public version of \samurai,\  which does not contain the extended 
decomposition - hence, implying a lighter reduction, with fewer
coefficients involved.

We remark that, within the integrand reduction algorithm, it is
possible to benefit immediately from the presence of powers of $q^2$
in the numerators, without any algebraic cost: the contribution of those
terms is automatically taken into account by the integrand
reconstruction of the subdiagrams (because they give no contribution
on the corresponding massless cut). On the contrary, within a tensor
reduction algorithm, these terms would cancel only
after the algebraic manipulation of the integrand.

\medskip

The numerical  values of the one-loop amplitudes of the
processes~(\ref{Eq:Pproc1})  in a non-exceptional phase space point 
are collected in~\ref{App:Bench}. 
The values of the double and the single poles conform to 
 the universal singular behavior  of  dimensionally regulated 
one-loop amplitudes~\cite{Giele:1991vf, Kunszt:1994np, Catani:1996jh,  Catani:1996vz,  Catani:2000ef}.
After appropriate crossing to the $H \to$ 4-parton decay kinematics,
we compared our results with the ones presented in Table~I of
Ref.~\cite{Ellis:2005qe}, finding excellent agreement. 
Furthermore, converting our results for the $Hjj$-production
channels from DRED to the 't Hooft-Veltman scheme, we
are in perfect agreement with the most recent version of MCFM (v6.4).


\begin{figure}[htb]
\begin{center}
\includegraphics[width=7.5cm]{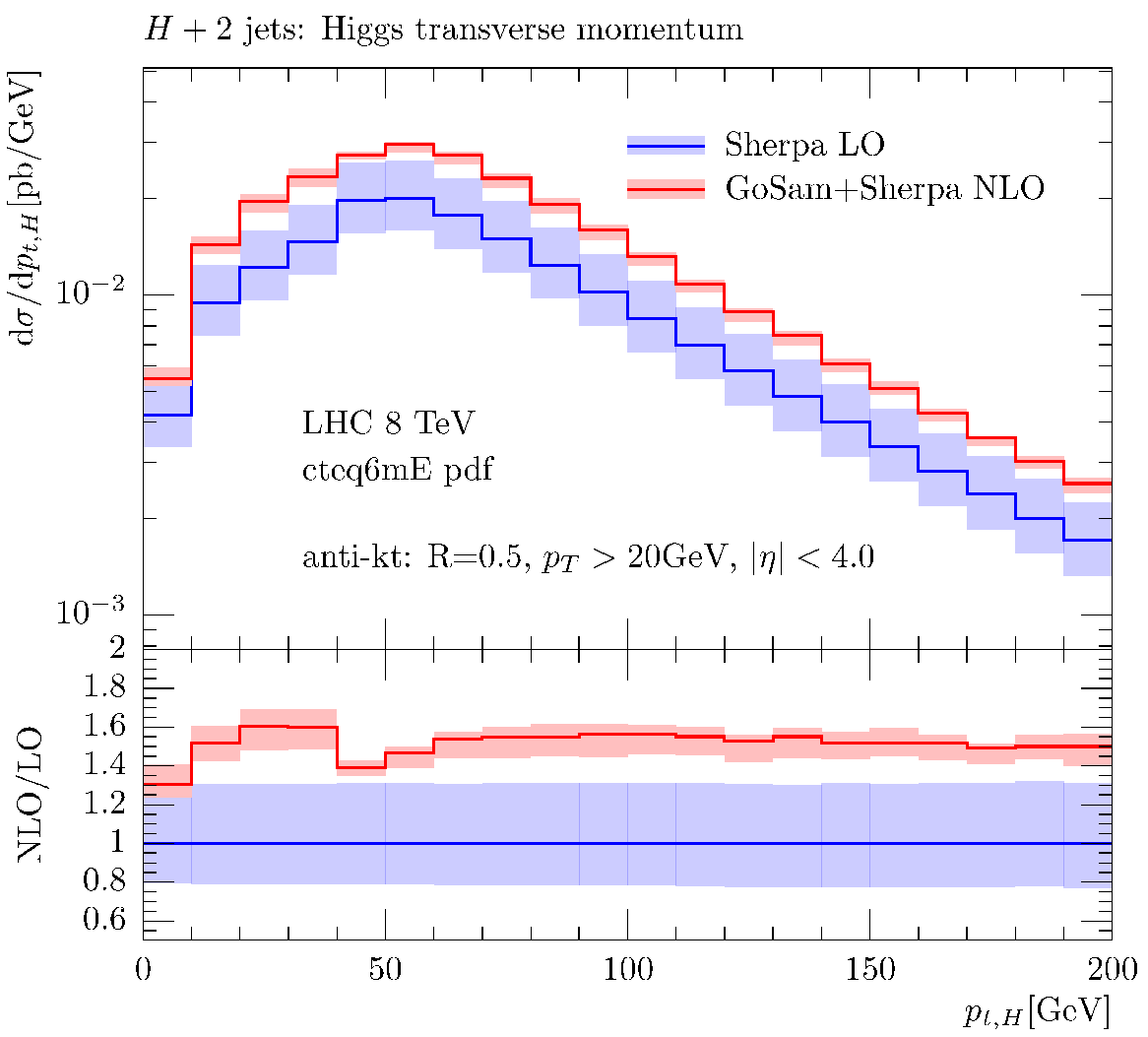} 
\caption{Transverse momentum $p_T$  of the Higgs boson. } \label{histo_PTh0}
\end{center}
\end{figure}

\begin{figure}[htb]
\begin{center}
\includegraphics[width=7.5cm]{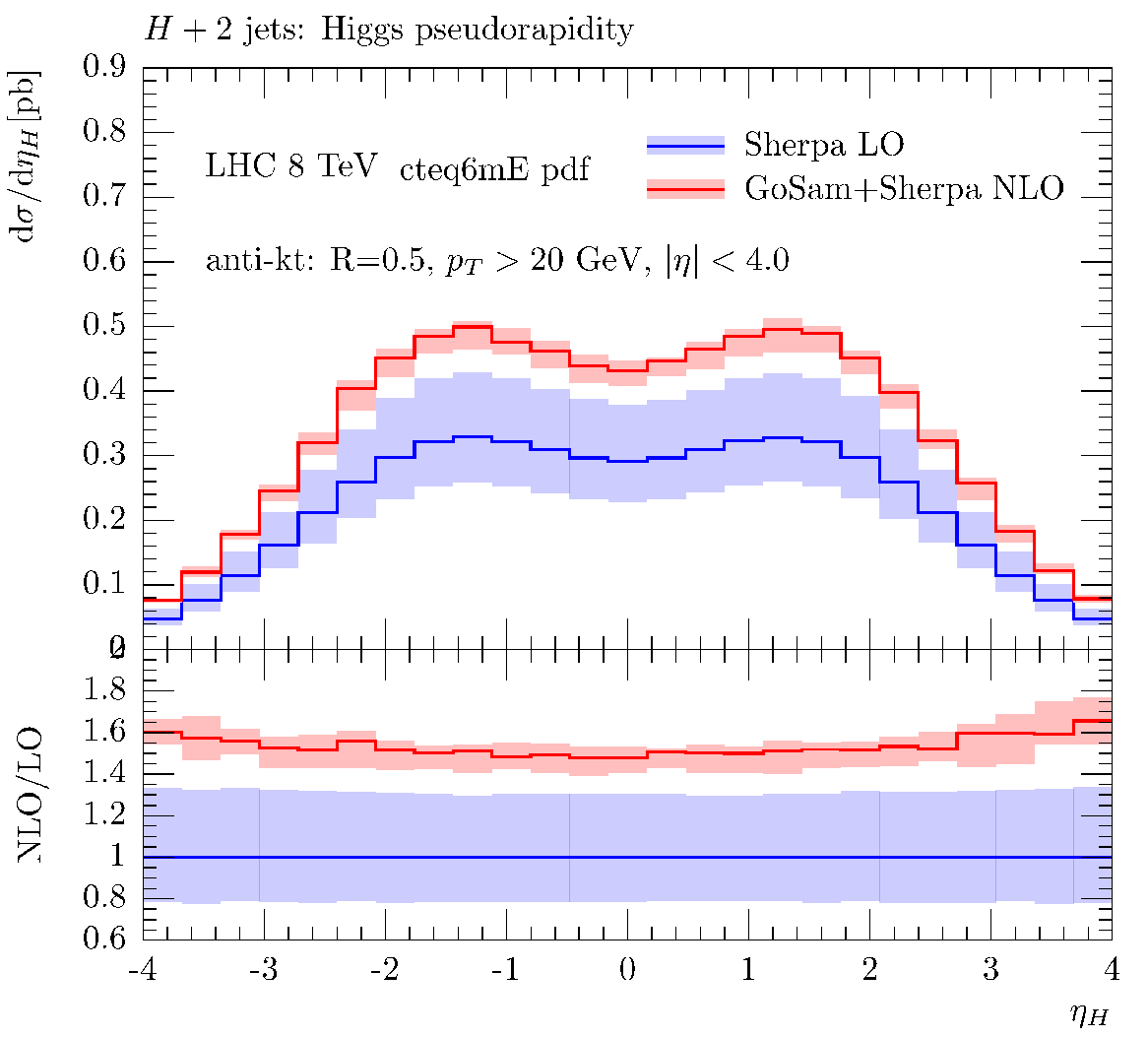} 
\caption{Pseudorapidity $\eta$ of the Higgs boson. } \label{histo_Etah0}
\end{center}
\end{figure}

\begin{figure}[htb]
\begin{center}
\includegraphics[width=7.5cm]{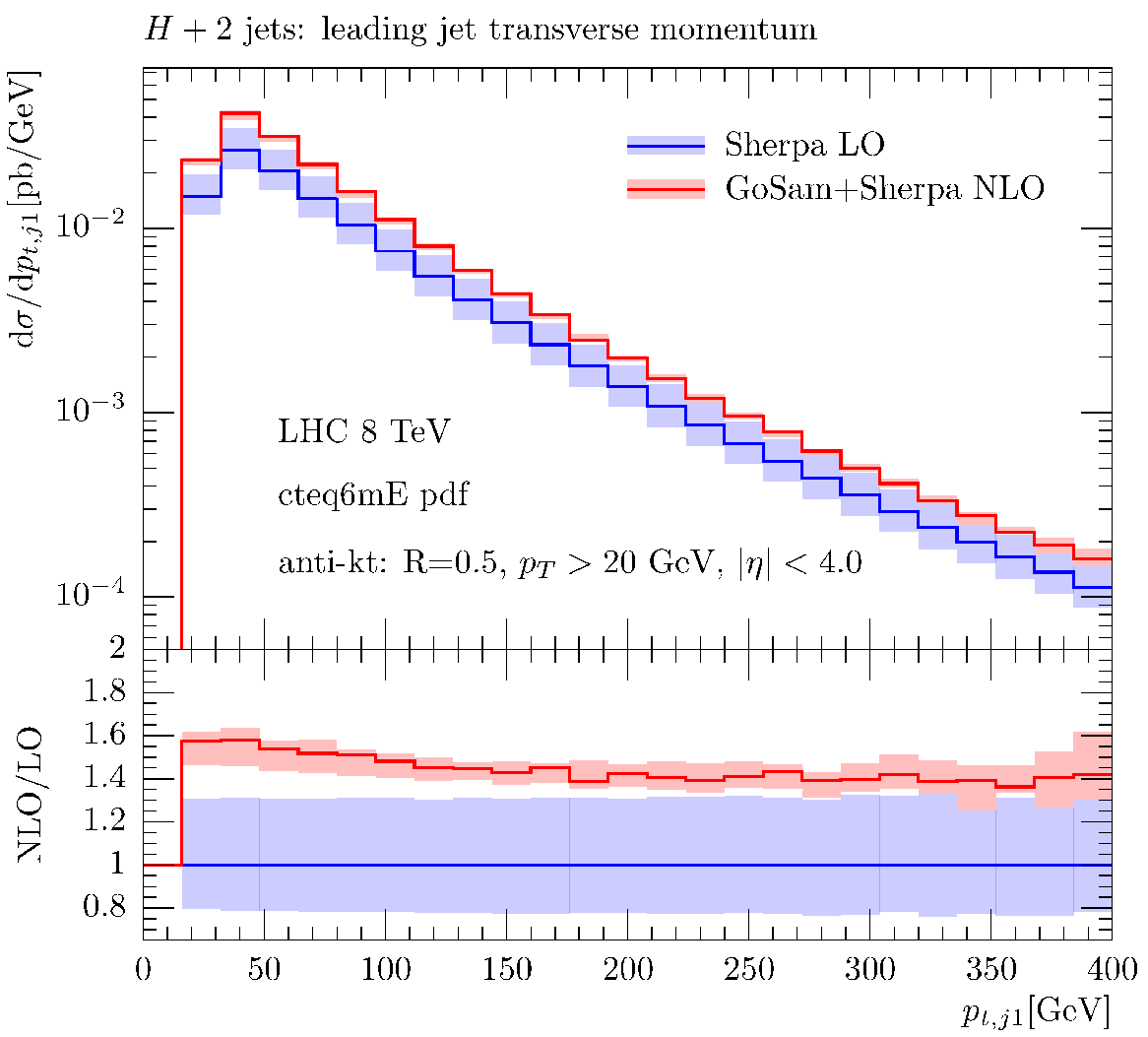}
\caption{Transverse momentum $p_T$ of the first jet.}  \label{histo_jets_jet_1_1_pt_1}
\end{center}
\end{figure}

\begin{figure}[htb]
\begin{center}
\includegraphics[width=7.5cm]{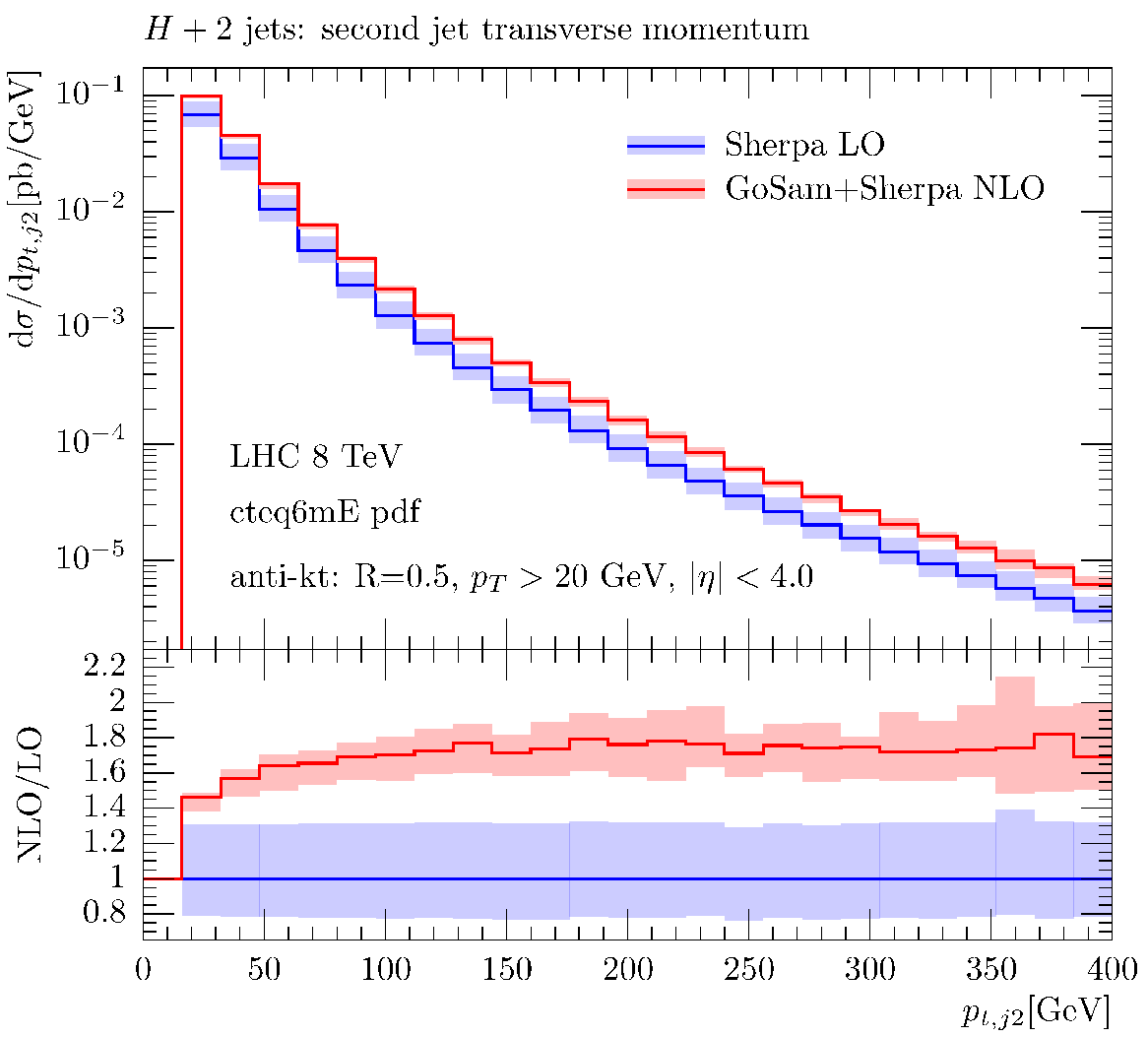}
\caption{Transverse momentum $p_T$ of the second jet. }  \label{histo_jets_jet_1_1_pt_2}
\end{center}
\end{figure}

\begin{figure}[htb]
\begin{center}
\includegraphics[width=7.5cm]{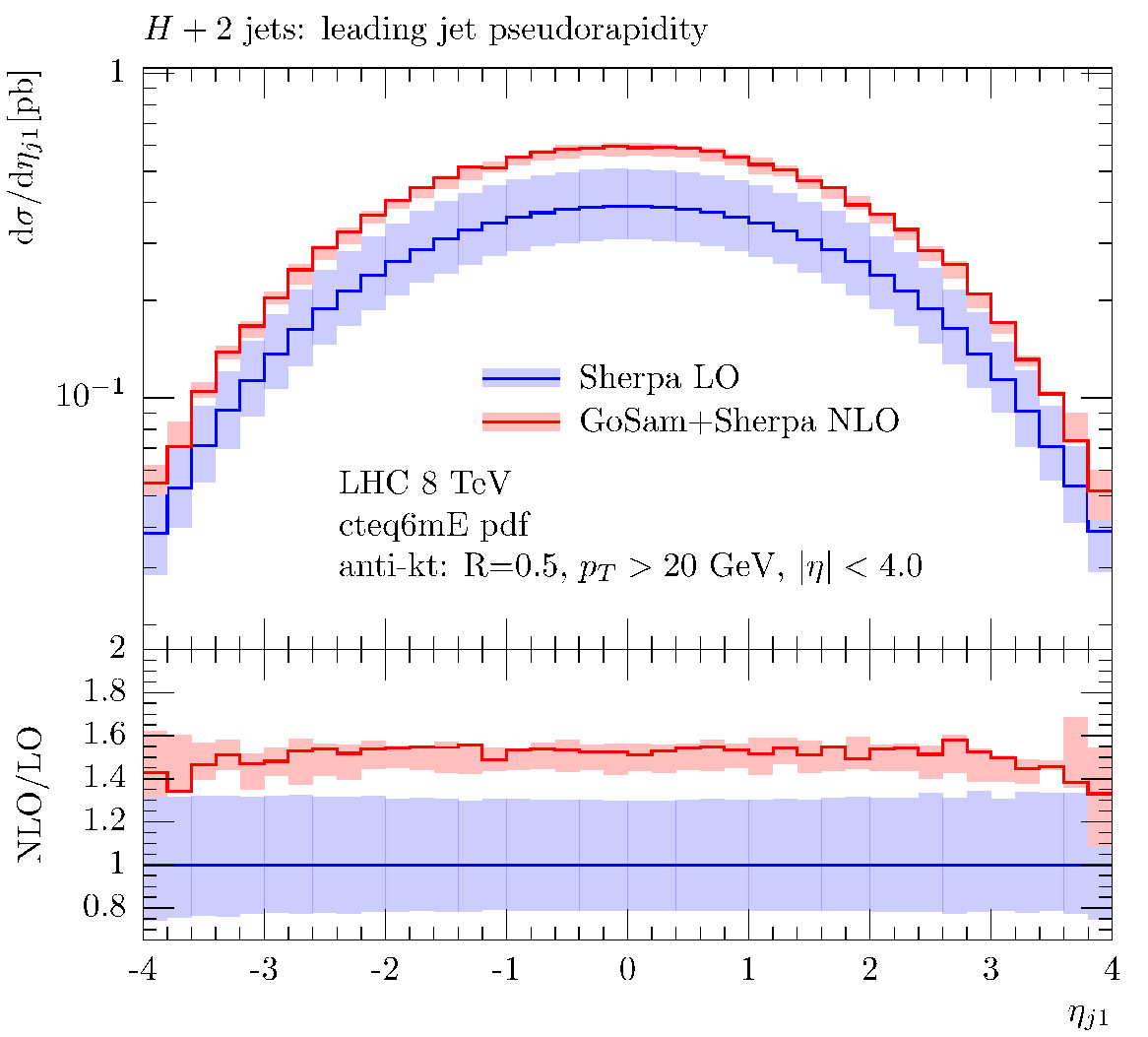}
\caption{Pseudorapidity $\eta$ of the first jet. }  \label{histo_jets_jet_1_1_eta_1}
\end{center}
\end{figure}

\begin{figure}[htb]
\begin{center}
\includegraphics[width=7.5cm]{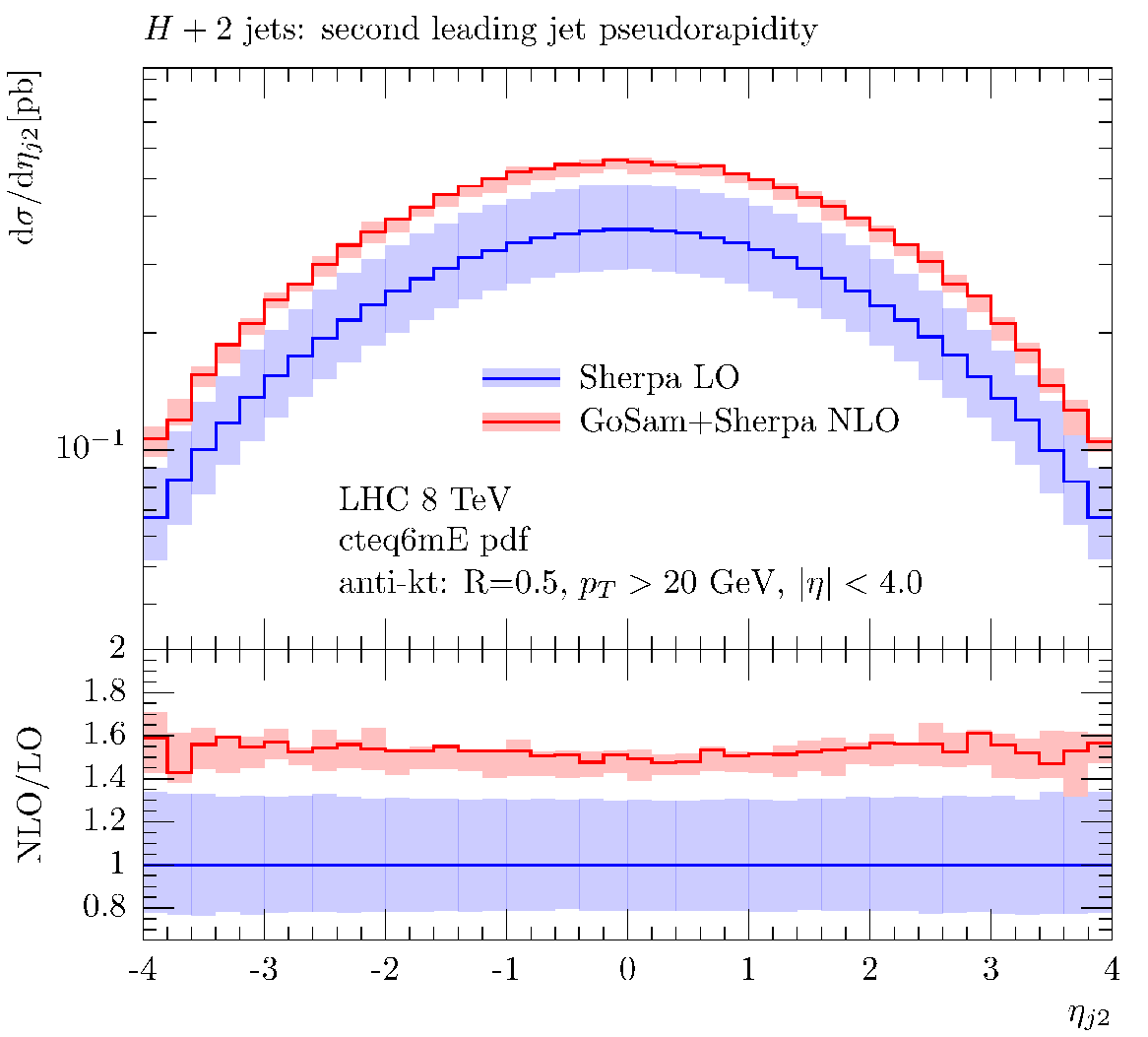}
\caption{Pseudorapidity $\eta$ of the second jet. } \label{histo_jets_jet_1_1_eta_2}
\end{center}
\end{figure}

\section{Numerical results for $pp \to Hjj$}
\label{sec:res}

In this section we present a selection of phenomenological results 
for proton-proton collisions at the LHC at $\sqrt{s}= 8 $ \TeV, as a
sample of the results that can be easily obtained with 
the \Gosam-{\Sherpa} automated setup~\cite{Cullen:2011ac,Gleisberg:2008ta,LST, packagesURL}.
 A more complete analysis of Higgs production in gluon fusion, which
 merges several multiplicities~\cite{Hoeche:2012yf} 
 and employs the code for the virtual matrix elements of $Hjj$
 presented here, is going to be discussed in \cite{sherpa}.
 
The results  shown in this section are obtained using the parameters listed below:
\begin{align}
&  M_H = 125 \gev \, ,     & & G_F = 1.16639 \cdot 10^{-5} \gev^{-2}  \, ,   \nonumber \\
& \alpha^{\mbox{\tiny LO}}_s(M_Z)= 0.129783 \, ,& & \alpha^{\mbox{\tiny NLO}}_s(M_Z)= 0.117981 \, , \nonumber \\
& v^2 = \frac{1}{\sqrt{2} G_F} \, .& & 
\label{Eq:InputP}
\end{align}
We use the CTEQ6L1 and CTEQ6mE~\cite{Pumplin:2002vw} parton distribution functions (PDF) for the LO and NLO, respectively. The value of the strong coupling at the scale $\mu$ 
is taken  from  the PDF set starting  from the  initial values in Eq.~(\ref{Eq:InputP}). The jets are clustered by using the anti-$k_T$ algorithm provided by the
\FastJet\ package~\cite{Cacciari:2005hq,Cacciari:2008gp,Cacciari:2011ma} with the following setup:
\beq
p_{t,j} \ge 20  \gev, \quad |\eta_j| \le 4.0, \quad  R = 0.5 \, .
\eeq
The Higgs boson is treated as a stable on-shell particle, without including any decay mode. 
To fix the factorization and the renormalization scale we define
\bea
\hat{H}_{t} = \sqrt{M_H^2+p_{t,H}^2}+\sum_{j} p_{t,j}  \, ,
\eea
where $p_{t,H}$ and $p_{t,j}$ are the transverse momenta of the Higgs boson and the jets.
The nominal value for the two scales is defined as
\bea
\mu = \mu_R = \mu_F = \hat{H}_{t}\, ,
\label{Eq:MUvar}
\eea
whereas theoretical uncertainties are assessed by varying simultaneously the factorization and renormalization scales in the range
\beq
\frac{1}{2}\hat{H}_{t} <  \mu  < 2 \hat{H}_{t}\,.
\eeq
The error is estimated by taking the envelope of the resulting distributions at the different scales.

\subsection{Results}
\label{sec:dist}
Within our framework, we find 
the following total cross sections for the process $p p \to H j j$ in gluon fusion:
$$\sigma_{\rm LO}[{\rm pb}] = 1.90^{+0.58}_{-0.41} \, ,$$
$$\sigma_{\rm NLO}[{\rm pb}]= 2.90^{+0.05}_{-0.20} \, ,$$
where the error is obtained by varying the renormalization and
factorization scales as given in Eq.~(\ref{Eq:MUvar}). 
The LO distributions have been computed using $2.5 \times 10^7$ phase
space points, whereas all NLO distributions have been obtained using
$4.0 \times 10^6$ phase space points for the Born and the virtual corrections and $5.0 \times 10^8$ points for the real radiation for each scale.

In Figs.~\ref{histo_PTh0} and~\ref{histo_Etah0}, we present the distribution of the transverse momentum $p_T$ of the Higgs boson and its pseudorapidity $\eta$, respectively. Both of them show a
K-factor between the LO and the NLO distribution of about $~1.5-1.6$, which is almost flat over a large fraction of kinematical range. Furthermore both plots show a decrease of the scale uncertainty
of about $50\%$.
Figures~\ref{histo_jets_jet_1_1_pt_1} and~\ref{histo_jets_jet_1_1_pt_2} display the transverse momentum of the first and second jet, whereas their pseudorapidities are shown in
Figs.~\ref{histo_jets_jet_1_1_eta_1}
and~\ref{histo_jets_jet_1_1_eta_2}. The previous considerations are
also true for these latter plots. 
For the transverse momentum distributions, however, we
observe a slight change of shapes.
In the case of the leading jet, increasing the $p_T$,
the K-factor decreases from 1.6 to 1.4; 
while for the second leading jet, it increases from 1.4 to 1.6.


\section{Virtual corrections to $pp \to Hjjj$}
\label{Sec:Hjjj}

We explore the possibility of extending our
framework to the production of a Higgs boson plus three jets at NLO.

The independent partonic processes contributing to $Hjjj$ 
can be obtained by adding one extra gluon to the final state of the
processes listed in Eq.(\ref{Eq:Pproc1}). Accordingly, we 
generate the codes for the virtual corrections to the partonic
processes with a quark-pair in the final state, 
\bea
gg \to H q {\bar q} g \ , \quad
q {\bar q} \to H q {\bar q} g \ , \quad
q {\bar q}  \to H q' {\bar q'} g \  . \quad
\label{Eq:H3qqbar}
\eea
The missing channel $gg \to  H g g g$,
together with the phase space integration, will be discussed in a successive study.

\begin{figure}[t]
\begin{center}
\includegraphics[width=9.0cm]{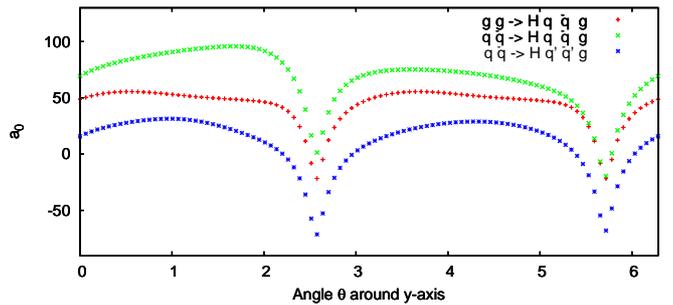} 
\caption{Finite-term of the virtual matrix-elements for 
$gg \to H q {\bar q} g $ (red),
$q {\bar q} \to H q {\bar q} g $ (green),
$q {\bar q}  \to H q' {\bar q'} g$ (blue).
} 
\label{Fig:Hjjj}
\end{center}
\end{figure}

We compute, for the first time, the virtual matrix elements for the three subprocesses listed above, and show
their results along a certain one-dimensional
curve in the space of final state momenta.
We take the initial partons to have momentum $p_1$ and  $p_2$, whose
3-momenta lie along the $z$-axis, and 
choose an arbitrary point for the final state momenta 
$\{p_3, p_4, p_5, p_6\}$. For simplicity, we start with the same phase space point used in the Appendix~D (see Table~\ref{Tab:ppsJ3}).
Then, we create new momentum configurations by rotating the final state
through an angle $\theta$ about the $y$-axis. 
Figure~\ref{Fig:Hjjj} displays the behavior of the finite part $a_0$ of
the individual $2 \to 4$ amplitudes defined as
\bea
\frac{  2 \mathfrak{Re} \left  \{ \mathcal{M}^{\mbox{\tiny tree-level} \ast } \mathcal{M}^{\mbox{\tiny one-loop}  }   \right  \}  }{
(4 \pi \alpha_s ) \left |  \mathcal{M}^{\mbox{\tiny tree-level}} \right |^2 
 }  
\equiv 
\frac{a_{-2}}{\epsilon^2} +  \frac{a_{-1}}{\epsilon} 
+ a_0   
\, ,
\label{Eq:AI}
\eea
when the final external momenta are rotated  from
$\theta=0$ to $\theta=2\pi$. The plots are obtained by sampling over $100$ points.

Numerical values for the one-loop amplitudes of the
processes listed in~(\ref{Eq:H3qqbar}) are collected in~\ref{App:BenchJ3}.

Also in this case we verify that 
the values of the double and the single poles conform to 
the universal singular behavior  of  dimensionally regulated 
one-loop amplitudes~\cite{Catani:2000ef}.

\section{Conclusions}
We presented the calculation of the associated
production of a Higgs boson and two jets, $pp \to H jj$, at NLO in
QCD, employing the infinite top-mass approximation.  

The results were obtained by using a fully automated framework for
fixed order NLO QCD calculations based on  the interplay of the packages 
\Gosam\ and  \Sherpa, interfaced through the BLHA standards.
We discussed the technical aspects of the computation,
and showed the numerical impact of the radiative corrections on
the distribution of the transverse momentum of the Higgs boson and its
pseudorapidity, as well as of the transverse momentum and pseudorapidity of the leading and second leading jet.
All plots show a K-factor between the LO and the NLO distributions of
about 1.5, over a large fraction of kinematical range, and a decrease of the scale uncertainty of about 50\%.

The evaluation of the virtual corrections constitutes an application of the 
$d$-dimensional integrand reduction to theories with higher dimensional operators.

Finally, as an initial step towards the evaluation of $pp \to H jjj$ at NLO,
we presented first results for the one-loop matrix elements
of the partonic processes with a quark-pair in the final state.

\section*{Acknowledgements}
We thank the \Sherpa\ collaboration for encouraging and stimulating discussions and feedback on the manuscript.
We also would like to thank Thomas Hahn for his technical support while structuring 
the computing resources needed by our codes, and Joscha Reichel for 
feedback on the extended-rank version of \samurai. 

The work of  H.v.D., G.L., P.M., and T.P. was  supported by the Alexander von
Humboldt Foundation, in the framework of the Sofja Kovaleskaja Award Project
``Advanced Mathematical Methods for Particle Physics'', endowed by the German
Federal Ministry of Education and Research. 

 G.O. was supported in part by the 
National Science Foundation under Grant PHY-1068550.

H.v.D. and G.L. thank the Center for Theoretical Physics of
New York City College of Technology for hospitality
during the final stages of this project.

The Feynman diagrams present in this paper are drawn using \Feynarts~\cite{Hahn:2000kx}.


\appendix

\section{Effective Higgs-gluon vertices}
\label{App:Feynman}

The operator $\mathcal{L}$ in Eq.~(\ref{Eq:EffL}) describes the 
gluon-Higgs interaction in the large top-mass limit and  leads to
the following set of Feynman rules:
 \begin{eqnarray}
&& \input{P3}  \nonumber   \\
&& \input{P4}  \nonumber   \\
&&  \input{P5} \nonumber 
\end{eqnarray}
where we define
\begin{align}
 \mathcal{F}^{\mu_1 \mu_2}_{c_1,c_2} & =  \delta_{c_1 c_2} \left ( 
p_1^{\mu_2} p_2^{\mu_1} - p_1\cdot p_2 \, g^{\mu_1 \mu_2}
\right ) \, ,  \nonumber \\[0.2ex]
 \mathcal{F}^{\mu_1 \mu_2 \mu_3}_{c_1,c_2, c_3}  & = f_{c_1c_2c_3} \big  [ \,
 g^{\mu_1 \mu_2} \,  (p_1^{\mu_3} - p_2^{\mu_3})  \nonumber \\
&+g^{\mu_2 \mu_3} \,    (p_2^{\mu_1} - p_3^{\mu_1})     \nonumber \\  
&+g^{\mu_3 \mu_1}  \,  (p_3^{\mu_2} - p_1^{\mu_2}) \,  \big  ]  \, ,  \nonumber  \\[0.2ex]
 \mathcal{F}^{\mu_1 \mu_2 \mu_3 \mu_4}_{c_1,c_2, c_3,c_4} 
 & =   f_{c_1 c_2 i} f_{c_3 c_4 i} [\, g^{\mu_1 \mu_4} g^{\mu_2 \mu_3} - g^{\mu_1 \mu_3} g^{\mu_2 \mu_4} \, ]  \nonumber \\
& +  f_{c_1 c_3 i} f_{c_2 c_4 i} [ \, g^{\mu_1 \mu_4} g^{\mu_2 \mu_3} - g^{\mu_1 \mu_2} g^{\mu_3 \mu_4} \, ] \nonumber \\
& +   f_{c_1 c_4 i} f_{c_2 c_3 i} [\, g^{\mu_1 \mu_3} g^{\mu_2 \mu_4} - g^{\mu_1 \mu_2} g^{\mu_3 \mu_4} \,  ] . 
\label{Eq:FFactors}
\end{align}
In Eq.~(\ref{Eq:FFactors}) sum over repeated indices is understood.

\section{Higher-rank integrands}
\label{App:Q2}

Higher-rank integrands, i.e. integrands where the  powers of loop momenta
in the  numerator is higher
than the numbers of denominators, are
present in diagrams with a Higgs boson coupled to a purely
gluonic loop involving only three-gluon vertices.  The generic numerator
$\Gamma^{ \varepsilon_1  \cdots  \varepsilon_n }$  of a   $(n+1)$-denominator 
one-loop diagram can be written as
\bea
{\Gamma}^{ \varepsilon_1  \cdots  \varepsilon_n }  &\equiv& \mathcal{F}_{\mu_1 \mu_2}   \; 
\mathcal{G}^{\mu_1 \mu_2  \varepsilon_1  \cdots  \varepsilon_n } \, , 
\label{Eq:MainQ2}
\eea
where $\mathcal{F}_{\mu_1 \mu_2} $ is the $Hgg$ vertex defined in Eq.~(\ref{Eq:FFactors}),
and $\mathcal{G}^{\mu_1 \mu_2  \varepsilon_1  \cdots  \varepsilon_n }$ is the numerator of 
an $(n+2)$-gluon tree-level diagram, which can be represented by
\begin{eqnarray}
\input{WW} \input{PP}\input{GG} \nonumber 
\end{eqnarray}
We are interested in the leading behaviour in $q$ of
$\Gamma^{ \varepsilon_1  \cdots  \varepsilon_n } $, and we
want to show that the highest-rank terms, with rank $r=n+2$,
are proportional to the loop momentum squared, $q^2$.
In order to show it, we neglect all external momenta 
and all the terms proportional  to $q^2$.

From Eq.~(\ref{Eq:FFactors}), one trivially has
\bea
\mathcal{F}_{\mu_1 \mu_2}   =  q_{\mu_1}  \,  q_{\mu_2} + \mathcal{O}( q^2)  \; ,
\label{Eq:AppQ2P}
\eea
while the  generic tensor structure of $\mathcal{G}^{\mu_1 \mu_2  \varepsilon_1  \cdots  \varepsilon_n } $  is 
\begin{eqnarray}
\mathcal{G}^{\mu_1 \mu_2  \varepsilon_1  \cdots  \varepsilon_n }  &=&  q^{\mu_1}  \mathcal{T}_1^{\mu_2  \varepsilon_1  \cdots  \varepsilon_n }+
q^{\mu_2}  \mathcal{T}_2^{\mu_1  \varepsilon_1  \cdots  \varepsilon_n
}  + \nonumber  \\
& & + \  g^{\mu_1 \mu_2}  \mathcal{T}_g^{  \varepsilon_1  \cdots  \varepsilon_n }  + \mathcal{O}(q^2) \, ,
\label{Eq:AppQ2G}
\end{eqnarray}
where ${\cal T}_1, {\cal T}_2$, and ${\cal T}_g$ are tensors which may depend on $q$ as well.
Indeed Eq.~(\ref{Eq:AppQ2G})   is fulfilled  for $n=0,1$,
\begin{eqnarray}
\mathcal{G}^{\mu_1 \mu_2} &=& g^{\mu_1 \mu_2} \nonumber \\
\mathcal{G}^{\mu_1 \mu_2  \varepsilon_1 }   &=&  g^{\mu_1  \varepsilon_1}  q^{\mu_2} + 
 g^{\mu_2  \varepsilon_1}  q^{\mu_1}  - 2 g^{\mu_2 \mu_1}  q^{ \varepsilon_1}  \; ,
\end{eqnarray}
while for $n>1$ it can be proven by induction over $n$ by using
\bea
\mathcal{G}^{\mu_1 \mu_2  \varepsilon_1  \cdots  \varepsilon_n }   = \mathcal{G}_\mu^{ \phantom{\mu}   \mu_2 \varepsilon_1  \cdots  \varepsilon_{n-1} } 
 \mathcal{G}^{ \mu_1 \mu   \varepsilon_n  }   \, ,
\eea
that is
\begin{eqnarray}
\input{GG1}  \input{GG2}  \input{GG3} \nonumber 
\end{eqnarray}
Combining Eq.~(\ref{Eq:AppQ2P}) and Eq.~(\ref{Eq:AppQ2G}), it is easy to realize that
each rank-$(n+2)$ term of an $n+1$-denominator diagram $\Gamma^{ \varepsilon_1  \cdots  \varepsilon_n }$ is
proportional to $q^2$. The factor $q^2$ simplifies  against one denominator
leading  to a rank $n$ numerator of an
$n$-denominator integrand.

\section{Benchmark points for $pp \to Hjj$}
\label{App:Bench}

	\begin{table*}[t]
		\centering
		\begin{tabular}{c c c c c}
		\hline
			particle & $E$& $p_x$ & $p_y$ & $p_z$ \\ 
		\hline
		$p_1$ & 250.00000000000000 &  0.0000000000000000 &  0.0000000000000000 &  250.00000000000000 \\
		$p_2$ & 250.00000000000000 &  0.0000000000000000 &  0.0000000000000000 & -250.00000000000000 \\
		$p_3$ & 143.67785106160801 &  51.663364918413812 & -22.547134012261804 &  42.905108772983255 \\
		$p_4$ & 190.20318863787611 & -153.36110830475005 & -108.23578590696623 & -30.702411577195452 \\
		$p_5$ & 166.11896030051594 &  101.69774338633616 &  130.78291991922802 & -12.202697195787838 \\
		\hline
		\end{tabular}
		\caption{Benchmark phase space point for Higgs plus two jets production}
		\label{jjhjjmom}
	\end{table*}

In this appendix we provide numerical  results for the renormalized virtual contributions
to the processes~(\ref{Eq:Pproc1}), in correspondence with the phase
space point in Table~\ref{jjhjjmom}.
The parameters can be read from Eqs.~(\ref{Eq:InputP}), while the renormalization and
factorization scales are set to the Higgs mass value.
The assignment of the momenta proceeds as follows 
 \begin{align}
  g(p_1) \, g(p_2) \, & \to \, H(p_3) \, g(p_4) \, g(p_5)  \, ,   \nonumber \\                 
  g(p_1) \, g(p_2) \, & \to \, H(p_3) \, q(p_4) \, \bar q(p_5)  \, , \nonumber \\
  q(p_1) \, \bar q(p_2) \,& \to \, H(p_3) \, q(p_4)   \, \bar q(p_5)  \, , \nonumber \\                
  q(p_1) \, \bar q(p_2) \,&  \to \, H(p_3) \, q'(p_4) \, \bar q'(p_5)   \, .
  \label{Eq:ProBench}
  \end{align}
The results are collected in Table~\ref{gghggres} and are computed using DRED.  In the second column of the  table 
we provide the LO squared amplitude, 
\bea
c_0  \equiv  \frac{ \left |  \mathcal{M}^{\mbox{\tiny tree-level}} \right |^2   }{(4 \pi \alpha_s)^2 \ci^2  } \, ,
\eea
and the coefficients $a_i$ defined in Eq.~(\ref{Eq:AI}).
As a check on the reconstruction of the renormalized poles, 
in the last column we show the values of 
$a_{-1}$ and $a_{-2}$ obtained by the universal singular behavior 
 of the dimensionally regularized 
one-loop amplitudes~\cite{Catani:2000ef}. 

\newcommand{\trule}{\rule[-1.5mm]{0mm}{6mm}}

\begin{table}[h!]
\begin{tabular}{l r r} \hline \hline \trule
 & $g g \to H g g$  &  \\  
                           \hline  \trule
			$c_0$	 & $0.1507218951429643\cdot10^{-3} $ & 		             \\
			$a_0$    & $ 59.8657965614009  $ 	     & 		             \\
			$a_{-1}$ & $-26.4694115468536  $  	     & $-26.46941154671207 $ \\
			$a_{-2}$ & $-12.00000000000001 $	     & $-12.00000000000000$ \\  
			\hline
			\hline \trule 
&  $g g \to H q \bar q$     &\\
   	                   \hline  \trule		
	                   $c_0$	 & $ 0.5677813961826772\cdot10^{-6} $ & 		      \\
			$a_0$    & $ 66.6635142370683 $ 	      & 		      \\
			$a_{-1}$ & $-16.5816633315627 $  	      & $-16.58166333155405 $ \\
			$a_{-2}$ & $-8.66666666666669 $		      & $-8.666666666666668$ \\
			\hline
			\hline \trule 
&  $q \bar q \to H q \bar q$    &\\
   	                   \hline  \trule	
	                   $c_0$	 & $ 0.1099527895267439\cdot 10^{-5}$ & 		     \\
			$a_0$    & $  88.2959834057198 $ 	      & 		     \\
			$a_{-1}$ & $ -10.9673755313443 $  	      & $-10.96737553134440$ \\
			$a_{-2}$ & $ -5.33333333333332 $	      & $-5.333333333333334$ \\
			\hline
			\hline \trule			
& $q \bar q \to   H q' \bar q' $  & \\
   	                   \hline  \trule	
    		      $c_0$	  & $ 0.1011096724203529 \cdot10^{-6}$ & 		      \\
		$a_0$	  & $  33.9521626734153 $  	       & 		      \\
		$a_{-1}$  & $ -13.8649292834138 $ 	       & $-13.86492928341388$ \\
		$a_{-2}$  & $ -5.33333333333334 $ 	       & $-5.333333333333334$ \\
			\hline
			\hline 
\end{tabular}
\caption{Numerical results for the processes listed in Eq.~(\ref{Eq:ProBench})}
\label{gghggres}
\end{table}

\section{Benchmark points for $pp \to Hjjj$}
\label{App:BenchJ3}

\begin{table}[h!]
\begin{tabular}{l r r} \hline \hline \trule
&  $g g \to H q \bar q g$     &\\
   	                   \hline  \trule		
	                   $b_0$	 &  $0.6309159660038877\cdot10^{-4} $ & 		      \\
			$a_0$    & $48.68424097859422 $      & 		      \\
			$a_{-1}$ &  $	 -36.08277727147958$      & $ -36.08277728199094  $   \\
			$a_{-2}$ & 	$-11.66666666667209 $	      &$-11.66666666666667 $ \\
			\hline
			\hline \trule 
&  $q \bar q \to H   q \bar q g$    &\\
   	                   \hline  \trule	
	                       $b_0$	  & $  0.3609139855530763\cdot10^{-4} $ & 		      \\
		$a_0$	  & 	$ 69.32351140490162  $     & 		      \\
		$a_{-1}$  & 	  $ -29.98862932963380 $    & $ -29.98862932963629 $    \\
		$a_{-2}$  & 	  $  -8.333333333333339$    &  $-8.333333333333334$ \\
			\hline
			\hline \trule			
& $q \bar q \to   H q' \bar q' g$  & \\
   	                   \hline  \trule	
    		      $b_0$	 &  $0.2687990772405433\cdot10^{-5} $ & 		     \\
			$a_0$    &   $15.79262767177915$  & 		     \\
			$a_{-1}$ &  $-32.35320587070861$      & $-32.35320587073038$ \\
			$a_{-2}$ &$ -8.333333333333398 $&$ -8.333333333333332$    \\
			\hline
			\hline 
\end{tabular}
\caption{Numerical results for the processes listed in Eq.~(\ref{Eq:ProBenchJ3})}
\label{Tab:resJ3}
\end{table}

	\begin{table*}[ht]
		\centering
		\begin{tabular}{c c c c c}
		\hline
			particle & $E$& $p_x$ & $p_y$ & $p_z$ \\ 
		\hline
		$p_1$ & 250.00000000000000 &  0.0000000000000000 &  0.0000000000000000 &  250.00000000000000 \\
		$p_2$ & 250.00000000000000 &  0.0000000000000000 &  0.0000000000000000 & -250.00000000000000 \\
		$p_3$ &   	131.06896655823209   &     27.707264814722667   &    -13.235482900394146    &    24.722529472591685   \\
		$p_4$ &  	  164.74420140597425  &     -129.37584098675183    &   -79.219260486951597  &     -64.240582451932028    \\
		$p_5$ &  	 117.02953632773803    &    54.480516624273569    &    97.990504664150677    &   -33.550658370629378  \\
		$p_6$ & 	87.157295708055642     &   47.188059547755266  &     -5.5357612768047906     &   73.068711349969661 	 \\
		\hline
		\end{tabular}
		\caption{Benchmark phase space point for Higgs plus three jets production}
		\label{Tab:ppsJ3}
	\end{table*}

In this appendix we collect first numerical  results for the renormalized virtual contributions to 
 \begin{align}
  g(p_1) \, g(p_2) \, & \to \, H(p_3) \, q(p_4) \, \bar q(p_5)  \,   g(p_6)  \, , \nonumber \\
  q(p_1) \, \bar q(p_2) \,& \to \, H(p_3) \, q(p_4)   \, \bar  q(p_5)  \,   g(p_6)  \, , \nonumber \\                
  q(p_1) \, \bar q(p_2) \,&  \to \, H(p_3) \, q'(p_4) \, \bar q'(p_5)  \,   g(p_6)   \, .
  \label{Eq:ProBenchJ3}
  \end{align}
The results, collected in Table~\ref{Tab:resJ3}, have been computed 
using the parameters in Eqs.~(\ref{Eq:InputP}), with the renormalization and
factorization scales set to the Higgs mass value, and choosing
the phase space point given in Table~\ref{Tab:ppsJ3}.
In particular, in the second column of Table~\ref{Tab:resJ3}, we provide
the quantity $b_0$,
\bea
b_0  \equiv  \frac{ \left |  \mathcal{M}^{\mbox{\tiny tree-level}} \right |^2   }{(4 \pi \alpha_s)^3 \ci^2  } \, ,
\eea
and the coefficients $a_i$ defined in Eq.~(\ref{Eq:AI}).
In the third column we show the values of $a_{-1}$ and $a_{-2}$
obtained from the universal singular behavior of one-loop amplitudes.



\bibliographystyle{utphys} 
\bibliography{references.bib}

\end{document}

%% file: P3.tex
\unitlength=0.22bp%
\begin{feynartspicture}(300,300)(1,1)
\FADiagram{}
\FAProp(0.,10.)(11.,10.)(0.,){/ScalarDash}{0}
\FALabel(5.5,8.93)[t]{$H$}
\FAProp(20.,15.)(11.,10.)(0.,){/Cycles}{0}
\FALabel(15.2273,13.3749)[br]{$g_1$}
\FAProp(20.,5.)(11.,10.)(0.,){/Cycles}{0}
\FALabel(14.8873,6.01315)[tr]{$g_2$}
\FALabel(36.3,10.0)[]{$= -i g_{\mbox{\tiny eff}}    \mathcal{F}^{\mu_1 \mu_2}_{c_1,c_2}$}
\FAVert(11.,10.){0}
\end{feynartspicture}
%

%% file: P4.tex
\unitlength=0.22bp%
\begin{feynartspicture}(300,300)(1,1)
\FADiagram{}
\FAProp(0.,15.)(10.,10.)(0.,){/ScalarDash}{0}
\FALabel(4.78682,11.5936)[tr]{$H$}
\FAProp(0.,5.)(10.,10.)(0.,){/Cycles}{0}
\FALabel(5.21318,6.59364)[tl]{$g_3$}
\FAProp(20.,15.)(10.,10.)(0.,){/Cycles}{0}
\FALabel(14.7868,13.4064)[br]{$g_1$}
\FAProp(20.,5.)(10.,10.)(0.,){/Cycles}{0}
\FALabel(15.2132,8.40636)[bl]{$g_2$}
\FALabel(37.7,10.0)[]{$= g_{\mbox{\tiny eff}}  g_s   \,  \mathcal{F}^{\mu_1 \mu_2 \mu_3}_{c_1,c_2, c_3} $}
\FAVert(10.,10.){0}
\end{feynartspicture}
%

%% file: P5.tex
\unitlength=0.22bp%
\begin{feynartspicture}(300,300)(1,1)
\FADiagram{}
\FAProp(0.,10.)(11.,10.)(0.,){/ScalarDash}{0}
\FALabel(2.5,8.93)[t]{$H$}
\FAProp(20.,15.)(10.,10.)(0.,){/Cycles}{0}
\FALabel(14.7868,13.4064)[br]{$g_2$}
\FAProp(20.,5.)(10.,10.)(0.,){/Cycles}{0}
\FALabel(15.2132,8.40636)[bl]{$g_3$}
\FAProp(10.,10.)(10.,17.)(0.,){/Cycles}{0}
\FALabel(6.0,15.0)[tl]{$g_1$}
\FAProp(10.,10.)(10.,3.)(0.,){/Cycles}{0}
\FALabel(6.0,7.0)[tl]{$g_4$}
\FALabel(40.0,10.0)[]{$=i g_{\mbox{\tiny eff}}  g^2_s   \,  \mathcal{F}^{\mu_1 \mu_2 \mu_3 \mu_4}_{c_1,c_2, c_3,c_4} $}
\FAVert(10.,10.){0}
\end{feynartspicture}
%

%% file: WW.tex
\unitlength=0.26bp%
\begin{feynartspicture}(300,300)(1,1)
\FADiagram{}
%
\FAProp(0.,10.)(6.5,10.)(0.,){/ScalarDash}{0}
%
%
%
\FAProp(6.5,10.)(13.5,15.)(0.,){/Cycles}{0}
\FAProp(6.5,10.)(13.5,5.)(0.,){/Cycles}{0}
%
%
%
%
\FAVert(6.5,10.){0}
%
%
\FAProp(13.5,15.)(13.5,11.)(0.,){/Cycles}{0}
\FAProp(13.5,11.)(13.5,9.)(0.,){/Cycles}{0}
\FAProp(13.5,9.)(13.5,7.)(0.,){/GhostDash}{0}
\FAProp(13.5,7.)(13.5,5.)(0.,){/Cycles}{0}
\FAVert(13.5,5.){0}
\FAVert(13.5,11.){0}
\FAVert(13.5,15.){0}
\FAProp(13.5,11.)(20.5,11.)(0.,){/Cycles}{0}
\FALabel(17.,12.2803)[b]{$\varepsilon_2$}
\FAProp(13.5,15.)(20.5,15.)(0.,){/Cycles}{0}
\FALabel(17.,16.2803)[b]{$\varepsilon_1$}
\FAProp(13.5,5.)(20.5,5.)(0.,){/Cycles}{0}
\FALabel(17.,3.71969)[t]{$\varepsilon_n$}
\FALabel(23.,9.9)[]{$\equiv$}
\end{feynartspicture}

%% file: PP.tex
\unitlength=0.26bp%
\begin{feynartspicture}(300,300)(1,1)
\FADiagram{}
%
\FAProp(4.,10.)(10.5,10.)(0.,){/ScalarDash}{0}
%
%
%
\FAProp(10.5,10.)(17.5,15.)(0.,){/Cycles}{0}
\FALabel(14.,16.2803)[b]{$\mu_2$}
\FAProp(10.5,10.)(17.5,5.)(0.,){/Cycles}{0}
\FALabel(14.,3.71969)[t]{$\mu_1$}
\FAProp(16.3,12.)(16.3,8.)(1.,){/Straight}{-1}
\FALabel(18., 10. )[]{\small $q$}
%
%
%
\FAVert(10.5,10.){0}
%
\end{feynartspicture}

%% file: GG.tex
\unitlength=0.26bp%
\begin{feynartspicture}(300,300)(1,1)
\FADiagram{}
\FAProp(0.,15.)(6.5,15.)(0.,){/Cycles}{0}
\FALabel(3.59853,16.2803)[b]{$\mu_2$}
\FAProp(0.,5.)(6.5,5.)(0.,){/Cycles}{0}
\FALabel(3.59853,3.71969)[t]{$\mu_1$}
\FAProp(6.5,15.)(6.5,11.)(0.,){/Cycles}{0}
\FAProp(6.5,11.)(6.5,9.)(0.,){/Cycles}{0}
\FAProp(6.5,9.)(6.5,7.)(0.,){/GhostDash}{0}
\FAProp(6.5,7.)(6.5,5.)(0.,){/Cycles}{0}
%
%
\FAProp(6.5,11.)(13.5,11.)(0.,){/Cycles}{0}
\FALabel(10.,12.2803)[b]{$\varepsilon_2$}
\FAProp(6.5,15.)(13.5,15.)(0.,){/Cycles}{0}
\FALabel(10.,16.2803)[b]{$\varepsilon_1$}
\FAProp(6.5,5.)(13.5,5.)(0.,){/Cycles}{0}
\FALabel(10.,3.71969)[t]{$\varepsilon_n$}
\FAProp(1.5,12.)(1.5,8.)(-1.,){/Straight}{1}
\FALabel(0., 10. )[]{\small $q$}
\FAVert(6.5,5.){0}
\FAVert(6.5,11.){0}
\FAVert(6.5,15.){0}
\end{feynartspicture}

%% file: GG1.tex
\unitlength=0.26bp%
\begin{feynartspicture}(300,300)(1,1)
\FADiagram{}
\FAProp(0.0,15.)(0.0,10.)(0.,){/Cycles}{0} 
\FAProp(0.0,10.)(0.0,5.)(0.,){/Cycles}{0} 
\FAProp(13.5,15.)(13.5,10.)(0.,){/Cycles}{0} 
\FAProp(13.5,10.)(13.5,5.)(0.,){/Cycles}{0} 
\FAProp(5.5,10.)(5.5,15.)(0.,){/Cycles}{0} 
\FAProp(0.0,10.)(5.5,10.)(0.,){/Cycles}{0} 
\FAProp(5.5,10.)(7.8,10.)(0.,){/Cycles}{0} 
\FAProp(11.2,10.)(13.5,10.)(0.,){/Cycles}{0} 
\FAProp(7.8,10.0)(11.2,10.)(0.,){/GhostDash}{0} 
\FALabel(18.,9.9)[]{$\equiv$}
\FAVert(0.0,10.){0}
\FAVert(13.5,10.){0}
\FAVert(5.5,10.){0}
\FALabel(0.0,16.2803)[b]{$\varepsilon_1$}
\FALabel(5.5,16.2803)[b]{$\varepsilon_2$}
\FALabel(13.5,16.2803)[b]{$\varepsilon_n$}
\FAProp(4.75,6.5)(8.75,6.5)(0.,){/Straight}{1}
\FALabel(6.75, 4.5)[]{\small $q$}
\FALabel(0.0,3.71969)[t]{$\mu_2$}
\FALabel(13.5,3.71969)[t]{$\mu_1$}
\end{feynartspicture}

%% file: GG2.tex
\unitlength=0.26bp%
\begin{feynartspicture}(300,300)(1,1)
\FADiagram{}
\FAProp(0.0,15.)(0.0,10.)(0.,){/Cycles}{0} 
\FAProp(0.0,10.)(0.0,5.)(0.,){/Cycles}{0} 
\FAProp(13.5,15.)(13.5,10.)(0.,){/Cycles}{0} 
\FAProp(18.,10.)(13.5,10.)(0.,){/Cycles}{0} 
\FAProp(5.5,10.)(5.5,15.)(0.,){/Cycles}{0} 
\FAProp(0.0,10.)(5.5,10.)(0.,){/Cycles}{0} 
\FAProp(5.5,10.)(7.8,10.)(0.,){/Cycles}{0} 
\FAProp(11.2,10.)(13.5,10.)(0.,){/Cycles}{0} 
\FAProp(7.8,10.0)(11.2,10.)(0.,){/GhostDash}{0} 
\FAVert(0.0,10.){0}
\FAVert(13.5,10.){0}
\FAVert(5.5,10.){0}
\FALabel(0.0,16.2803)[b]{$\varepsilon_1$}
\FALabel(5.5,16.2803)[b]{$\varepsilon_2$}
\FALabel(13.5,16.2803)[b]{$\varepsilon_{n-1}$}
\FALabel(0.0,3.71969)[t]{$\mu_2$}
\FALabel(16.0,6.5)[b]{$\mu$}
%
%
\FAProp(7.8,6.5)(11.2,6.5)(0.,){/Straight}{1} 
\FALabel(9.5, 4.5)[]{\small $q$}
\end{feynartspicture}

%% file: GG3.tex
\unitlength=0.26bp%
\begin{feynartspicture}(300,300)(1,1)
\FADiagram{}
\FAProp(6.5,10.)(0.0,10.)(0.,){/Cycles}{0} 
\FAProp(6.5,10.)(6.5,15.)(0.,){/Cycles}{0} 
\FAProp(6.5,10.)(6.5,5.)(0.,){/Cycles}{0} 
\FAVert(6.5,10.){0}
\FALabel(6.5,16.2803)[b]{$\varepsilon_n$}
\FALabel(6.5,3.71969)[t]{$\mu_1$}
\FALabel(3.2,6.5)[b]{$\mu$}
%
\end{feynartspicture}